\documentclass[12pt]{article}
\makeatletter
\def\maxwidth{ %
  \ifdim\Gin@nat@width>\linewidth
    \linewidth
  \else
    \Gin@nat@width
  \fi
}
\makeatother

\RequirePackage{amsmath,amssymb}
\RequirePackage{amsthm}
\RequirePackage{natbib}
\RequirePackage[colorlinks,allcolors=blue]{hyperref}

\usepackage{fullpage}
\usepackage[nodisplayskipstretch]{setspace}

\usepackage{bbm,graphicx,bm} 
\usepackage{enumitem}
\usepackage{multirow}
\usepackage[all,import]{xy}
\usepackage{append ix}
\usepackage{comment}
\usepackage{color}
\usepackage{soul}
\usepackage{pdflscape}
\usepackage{subcaption}

\usepackage{booktabs}
\usepackage{longtable}
\usepackage{array}
\usepackage[table]{xcolor}
\usepackage{wrapfig}
\usepackage{float}
\usepackage{colortbl}
\usepackage{pdflscape}
\usepackage{tabu}
\usepackage{threeparttable}
\usepackage{threeparttablex}
\usepackage[normalem]{ulem}
\usepackage{makecell}
\usepackage{afterpage}
\usepackage{thmtools}
\usepackage[compact]{titlesec}

\usepackage{tikz}              
\usetikzlibrary{shapes,arrows,positioning,fit, shapes.geometric, patterns, patterns.meta, automata, calc,tikzmark,patterns,decorations,decorations.pathmorphing,decorations.text,decorations.pathreplacing,calligraphy, backgrounds, overlay-beamer-styles}

\theoremstyle{definition}

\newcommand{\bbone}{\ensuremath{\mathbbm{1}}}

\def\b1{\boldsymbol{1}}

\definecolor{purp}{RGB}{117, 112, 179}
\definecolor{orng}{RGB}{217, 95, 2}
\definecolor{grn}{RGB}{27,158,119}

\def\spacingset#1{\renewcommand{\baselinestretch}%
{#1}\small\normalsize} \spacingset{1}

\definecolor{RED}{RGB}{255,0,0}

\usepackage{soul}
\usepackage[utf8]{inputenc}

\begin{document}

\newcommand{\blind}{0}

\newcommand{\tit}{Discussion of ``Matrix Completion When Missing Is Not at Random and Its Applications in Causal Panel Data Models''}

\if0\blind

\title{\tit}

\date{}

\author{Eli Ben-Michael\thanks{Assistant Professor, Department of
    Statistics \& Data Science and Heinz College of Information
    Systems \& Public Policy, Carnegie Mellon University} \and
  \and Avi Feller\thanks{Associate Professor, Goldman School of Public
    Policy \& Department of Statistics, University of California, Berkeley}
}

\maketitle
\pagenumbering{gobble}
\setcounter{page}{1}
\pagenumbering{arabic}



\onehalfspacing
\section{Introduction}

Choi and Yuan (2025) (hereafter CY) propose a novel approach to applying matrix completion to the problem of estimating causal effects in panel data.
The key insight is that even in the presence of structured patterns of missing data---i.e. selection into treatment---matrix completion can be effective if the number of treated observations is small relative to the number of control observations.
We applaud the authors for their insightful and interesting paper.
Below, we situate their proposal as an example of a ``split-apply-combine'' strategy for panel data estimators (Section~\ref{sec:split-apply-combine}). We then discuss the issue of the statistical ``last mile problem''---the gap between theory and practice---and offer suggestions on how to partially address it (Section~\ref{sec:last_mile}). We conclude by considering the challenges of estimating the impacts of public policies using panel data and apply the approach to a study on the effect of right to carry laws on violent crime (Section~\ref{sec:policy_eval}). 
To do so, we extend CY's setup using the following notation.
Let $G_i \in \{1,\ldots,T,\infty\}$ indicate the treatment time for unit $i=1,\ldots,N$, $D_{it} = \bbone\{t \geq G_i\}$ be the treatment indicator for unit $i$ at time $t=1,\ldots,T$, and $Y_{it}(g)$ denote the potential outcome for unit $i$ at time $t$ if it were to be treated in period $g$.
Under the assumptions of no anticipation and no interference, the observed outcome $Y_{it}$ is related to the potential outcomes as $Y_{it} = Y_{it}(\infty) (1-D_{it}) + Y_{it}(G_i) D_{it}$ \citep{athey2018design}.

\section{The split-apply-combine strategy}
\label{sec:split-apply-combine}
CY's proposal uses a``split-apply-combine'' strategy that
``break[s] up a big problem into manageable pieces, operate[s] on each piece independently and then put[s] all the pieces back together'' \citep{wickham_split-apply-combine_2011}.
This framework is the backbone of many modern panel data estimators, including difference in differences and synthetic control approaches.

\subsection{Split-apply-combine panel data estimators}
\label{sec:did_etc}

\begin{figure}
  \begin{subfigure}{0.5\textwidth}
    \begin{center}
    \begin{tikzpicture}[node distance=1.2cm, >=stealth, thick]
      \node[] (left) {
        \includegraphics[height=0.08\textheight]{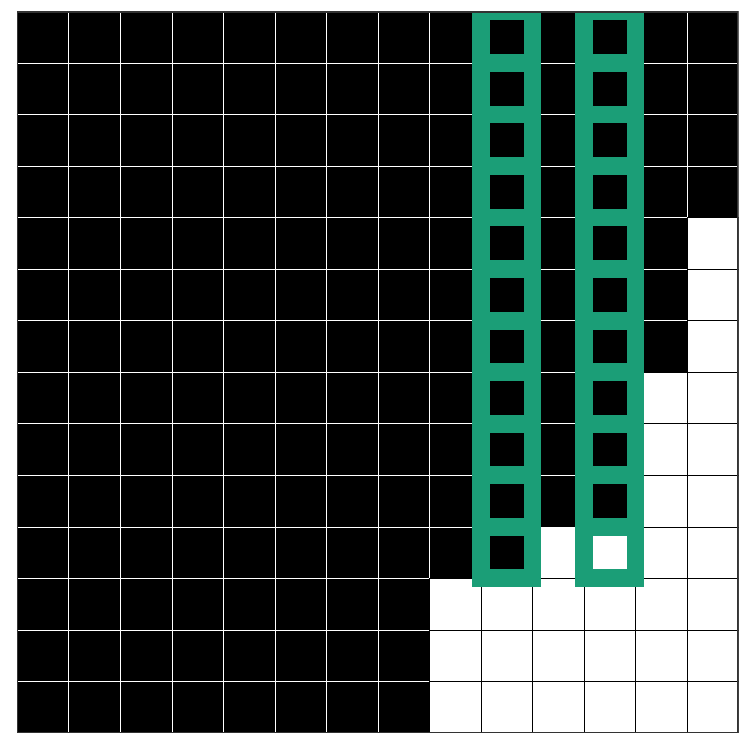}%
      };
      \node[rectangle,align=center, right=1.5cm of left, yshift=.8cm, draw=none] (mid1) {\includegraphics[height = 0.025\textheight]{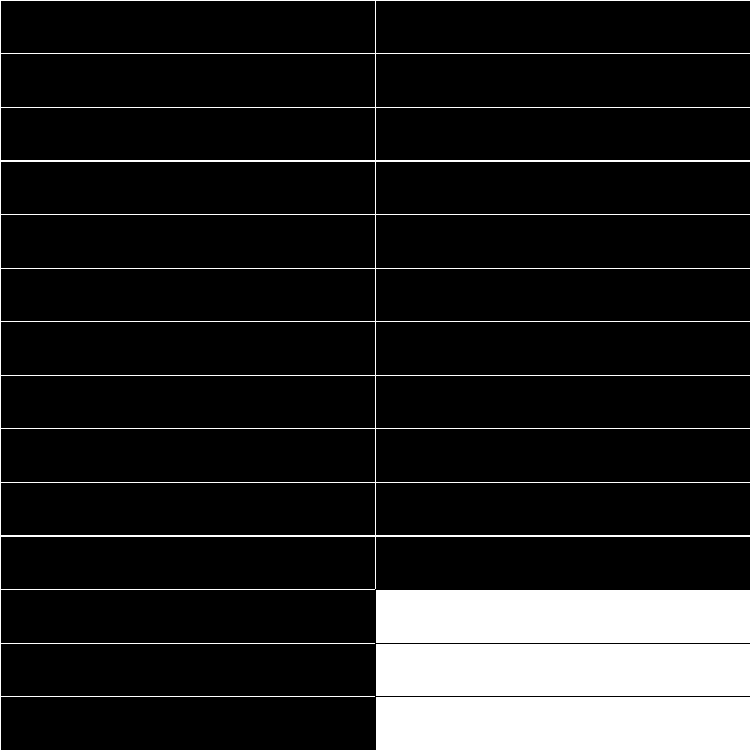}};
      \node[rectangle, align=center, right=1.5cm of left, yshift=0cm, draw=none] (mid2) {\includegraphics[height = 0.025\textheight]{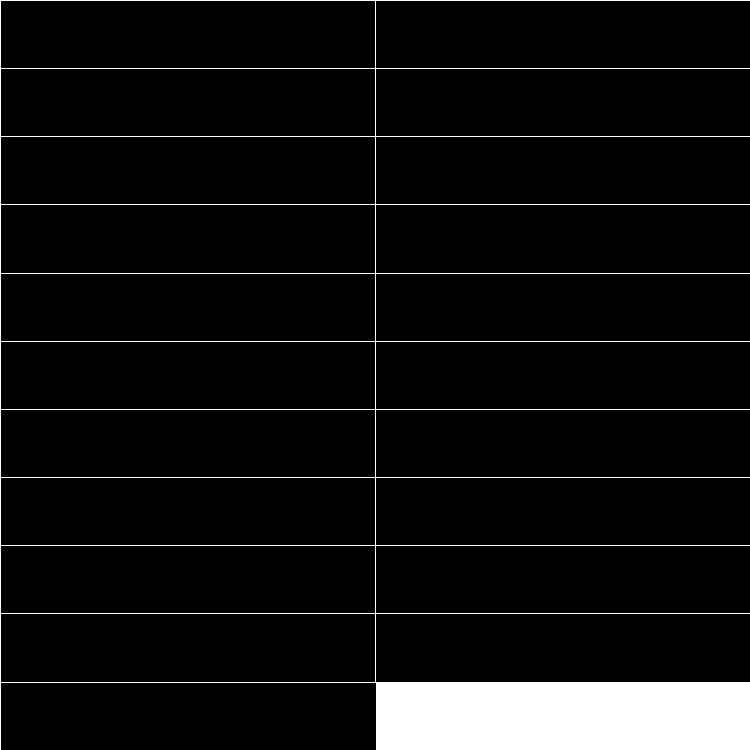}};
      \node[rectangle, align=center, right=1.5cm of left, yshift=-.8cm, draw=none] (mid3) {\includegraphics[height = 0.025\textheight]{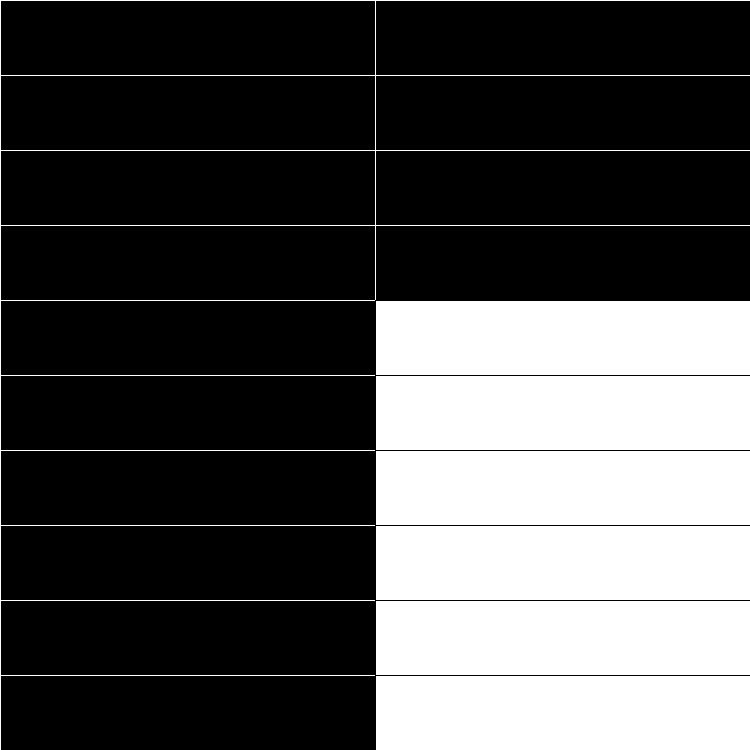}};
      \node[rectangle, align=center, right=1.5cm of mid2, draw=none] (right) {\Large \color{purp}$\hat{\tau}$};
      \draw[->, color=grn, line width=1.2pt] (left) -- (mid1);
      \draw[->, color=grn, line width=1.2pt] (left) -- (mid2);
      \draw[->, color=grn, line width=1.2pt] (left) -- (mid3);
      \draw[->, color=purp, line width=1.2pt] (mid1) -- (right);
      \draw[->, color=purp, line width=1.2pt] (mid2) -- (right);
      \draw[->, color=purp, line width=1.2pt] (mid3) -- (right);
      \begin{scope}[on background layer]
        \node[
        fit=(mid1)(mid3), 
        fill=orng!40, 
        rounded corners=6pt, 
        inner ysep=0pt, 
        inner xsep=0pt,
        draw=none
      ] (bgrect) {};
      \node[below=35pt of mid2.center, font=\bfseries, text=orng] (applylabel) {Apply};
      \end{scope}

      \node[below=3pt of applylabel.center, font=\scriptsize, text=orng] (splitlabel2) {2x2 DiD};
      \node[below=35pt of left.center, font=\bfseries, text=grn] (splitlabel) {Split};
      \node[below=3pt of splitlabel.center, font=\scriptsize, text=grn] {Treatment-time cohorts};
      \node[below=35pt of right.center, font=\bfseries, text=purp] (combinelabel) {Combine};
      \node[below=3pt of combinelabel.center, font=\scriptsize, text=purp] {Event time};
    \end{tikzpicture}
    
    \end{center}
    \caption{Difference in differences.}
    \label{fig:sac_did}
  \end{subfigure}
  \begin{subfigure}{0.5\textwidth}
    \begin{center}
    \begin{tikzpicture}[node distance=1.2cm, >=stealth, thick]
      \node[] (left) {
        \includegraphics[height=0.08\textheight]{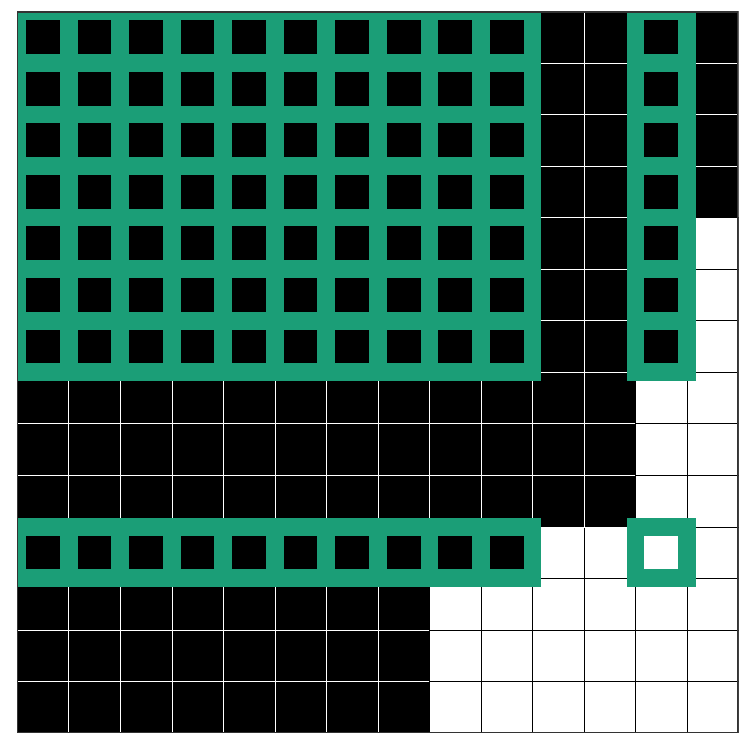}%
      };
      \node[rectangle,align=center, right=1.5cm of left, yshift=.8cm, draw=none] (mid1) {\includegraphics[height = 0.025\textheight]{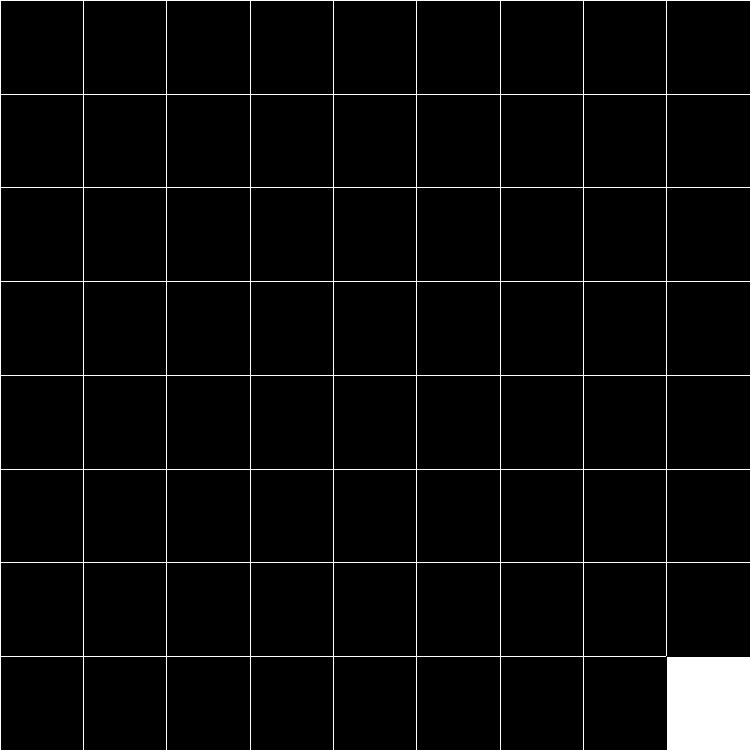}};
      \node[rectangle, align=center, right=1.5cm of left, yshift=0cm, draw=none] (mid2) {\includegraphics[height = 0.025\textheight]{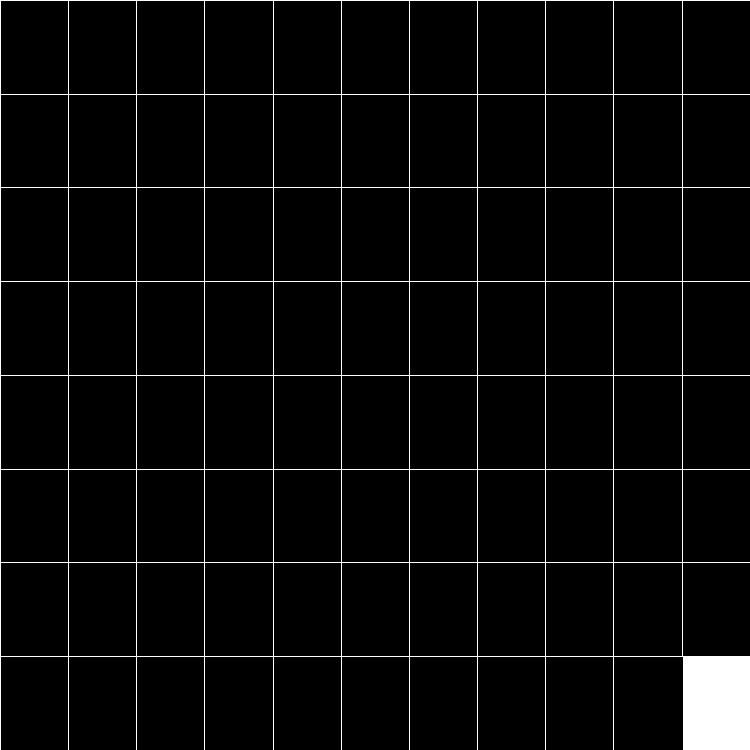}};
      \node[rectangle, align=center, right=1.5cm of left, yshift=-.8cm, draw=none] (mid3) {\includegraphics[height = 0.025\textheight]{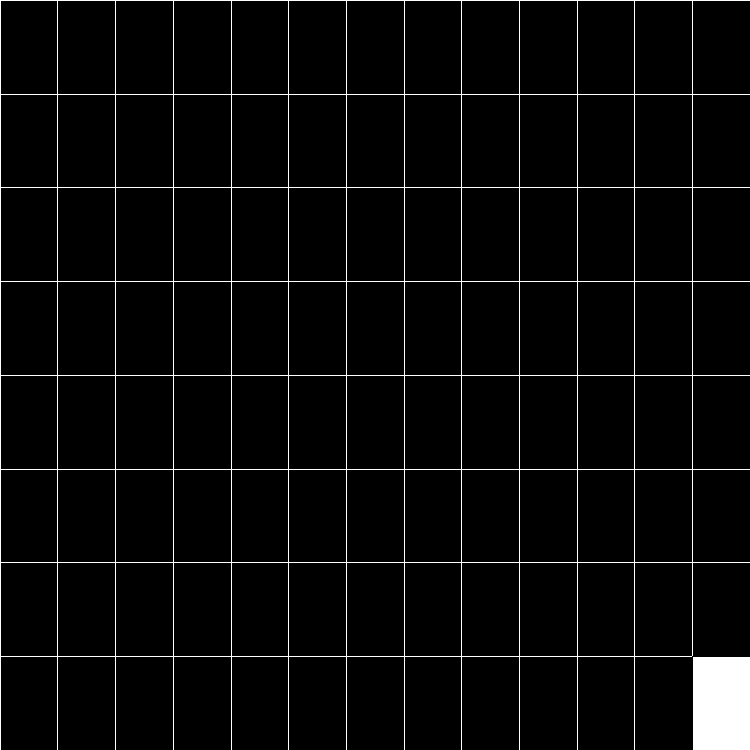}};
      \node[rectangle, align=center, right=1.5cm of mid2, draw=none] (right) {\Large \color{purp}$\hat{\tau}$};
      \draw[->, color=grn, line width=1.2pt] (left) -- (mid1);
      \draw[->, color=grn, line width=1.2pt] (left) -- (mid2);
      \draw[->, color=grn, line width=1.2pt] (left) -- (mid3);
      \draw[->, color=purp, line width=1.2pt] (mid1) -- (right);
      \draw[->, color=purp, line width=1.2pt] (mid2) -- (right);
      \draw[->, color=purp, line width=1.2pt] (mid3) -- (right);
      \begin{scope}[on background layer]
        \node[
        fit=(mid1)(mid3), 
        fill=orng!40, 
        rounded corners=6pt, 
        inner ysep=0pt, 
        inner xsep=0pt,
        draw=none
      ] (bgrect) {};
      \node[below=35pt of mid2.center, font=\bfseries, text=orng] (applylabel) {Apply};
      \end{scope}

      \node[below=3pt of applylabel.center, font=\scriptsize, text=orng] (splitlabel2) {Matrix completion};
      \node[below=35pt of left.center, font=\bfseries, text=grn] (splitlabel) {Split};
      \node[below=3pt of splitlabel.center, font=\scriptsize, text=grn] {Unit-time pairs};
      \node[below=35pt of right.center, font=\bfseries, text=purp] (combinelabel) {Combine};
      \node[below=3pt of combinelabel.center, font=\scriptsize, text=purp] {Calendar time};
    \end{tikzpicture}
    \end{center}
    \caption{Matrix completion in CY.}
    \label{fig:sac_mc}
  \end{subfigure}
  \caption{Split-apply-combine strategies for panel data estimators.}
  \label{fig:sac}
\end{figure}
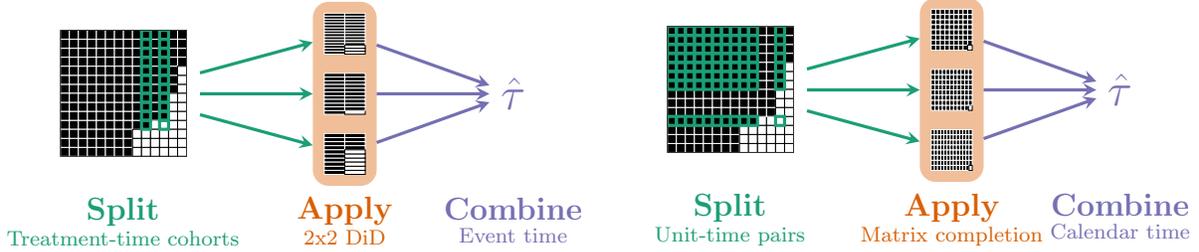

The split-apply-combine framework has been successful in transforming the way practitioners estimate causal effects with difference in differences (DiD) under staggered adoption.
Prior to recent methodological advances, common practice was to 
estimate a single two-way fixed effects (TWFE) regression model on the entire panel dataset: $Y_{it} = \alpha_i + \eta_t + \tau D_{it} + \varepsilon_{it}$,
where $\alpha_i$ and $\eta_t$ are unit and time fixed effects, respectively, 
and $\varepsilon_{it}$ is a mean-zero error term.
As CY note, this model estimates a single treatment effect $\tau$ and does not allow for treatment effect heterogeneity.
In the two-period, two-group case where no units are treated in the first period and some---but not all---units are treated in the second period, $\tau$ is the average treatment effect on the treated in the second period.
However, this does not necessarily hold in general: if there exist heterogeneous treatment effects by units and time, then $\tau$ may not even estimate a coherent average of effects \citep{goodman-bacon_did_2021,borusyak_revisiting_2024}.

Modern approaches to DiD address this challenge using the split-apply-combine strategy \citep{sun_did_2021,callaway_did_2021}.
The idea is to (i) split the panel into the possible two-period, two-group cases; (ii) estimate the effect $\tau$ from the TWFE model for each case; and (iii) combine estimates from the different cases to obtain a final estimate (see Figure~\ref{fig:sac_did}). This split-apply-combine strategy avoids so-called ``forbidden comparisons'' between already treated units and estimates a coherent combination of treatment effects.
Such split-apply-combine strategies are also found in synthetic control approaches with multiple treated units: (i) splitting into treated units or treatment-time cohorts; (ii) fitting a synthetic control for each treated unit or the average of the treatment-time cohort; and (iii) averaging the synthetic control estimates \citep[see, e.g.][]{abadie_penalized_2021, benmichael_multisynth_2022}.


CY thoughtfully employ the split-apply-combine strategy to matrix completion estimators.
The \textit{split} step first selects a focal time period and a set of units with the same treatment time before the focal period. It then subsets the matrix into the selected set of treated units and the full set of not-yet-treated units for the pre-treatment time periods and the focal time period.
The \textit{apply} step imputes the missing counterfactuals via nuclear-norm regularized matrix completion.
Finally, the \textit{combine} step averages the difference between the observed outcomes and the imputed counterfactual across treated units (see Figure~\ref{fig:sac_mc}).


\subsection{Combining before applying: lessons from weighting estimators}
\label{sec:combine_first}

CY consider two estimands: (i) the individual treatment effect (ITE) $\tau_{it} \equiv Y_{it}(G_i) - Y_{it}(\infty)$ for each treated unit, and (ii) the average treatment effect on the treated (ATT) at time $t_0$: $\tau^\text{cal}_{t_0} \equiv \frac{1}{N_{t_0}} \sum_{i: D_{it_0} = 1} \tau_{it_0}$, where $N_{t_0}$ is the number of treated units at time $t_0$, and where ``cal'' denotes calendar time (we return to this below).
While CY provide high probability bounds on the the imputation error for the structural component of the ITE,
we cannot have useful bounds for the estimation error of the ITE itself because of the idiosyncratic error term. This is \textit{the fundamental problem of causal inference}: it is not generally possible to fully impute individual-level counterfactuals.

This identifiability issue makes the ATT an attractive alternative when there are multiple treated units.  In this case
it is often useful to combine \emph{before} applying the imputation/treatment effect estimation step.
For instance, this is common with matching and weighting estimators in cross-sectional observational studies. Here the
\textit{split-apply-combine} strategy is to:
(i) split the data by each treated unit, (ii) find a set of matched or re-weighted control units
with a similar covariate profile to the treated unit and impute the counterfactual, and (iii) average the differences between the observed and imputed outcomes.

If the expected potential outcome conditional on the covariates has some amount of smoothness, this approach will be inefficient compared to the \emph{combine-apply} strategy: (i) average the treated units together and (ii) find a set of matched or re-weighted control units that has a similar covariate profile to the average treated unit, then estimate the difference in matched or re-weighted means.
Such ``fine balance'' \citep{zubizarreta_matching_2011} approaches have larger effective sample sizes.
Similar combine-apply strategies are also useful in synthetic control settings \citep[see, e.g.][]{Kreif2016, Robbins2017, benmichael_multisynth_2022}.
For the matrix completion estimator that CY propose, such a combine-apply strategy may also be effective when targeting the ATT.

\section{The statistical last mile problem}
\label{sec:last_mile}
The ``last mile problem'' refers to the challenges and costs associated with the final step of delivering goods or services, such as moving a package from a distribution center to its final destination.
Statistical methodology suffers from such a last mile problem:
we often develop estimators with good theoretical guarantees under ideal settings, but face hurdles in deploying these tools in practice.
Overcoming this last mile problem remains a central challenge for our field and, we believe, is an important complement to cutting-edge core methodological development, like CY.
In that spirit, we offer some possible directions for overcoming the last mile problem in deploying CY's proposal.


\paragraph{Combining effects by calendar time vs event time.}
CY target a \emph{calendar time} average treatment effect, $\tau_{t_0}^\text{cal}$, that averages across all treated units at a given calendar time $t_0$.
The calendar time average is difficult to interpret when there are treatment effect dynamics: it aggregates effects for units at different points in the treatment process, averaging together units that only just received treatment and that have received treatment for many periods.

Much applied work instead focuses on the \emph{event time} average, which averages treatment effects across all treated units at a given event time $G_i + k$: $\tau^\text{event}_k \equiv \frac{1}{N_k} \sum_{i: G_i < \infty} \tau_{i,k+G_i}$, where $N_k$ is the number of treated units that have observed outcomes $k$ periods after treatment (i.e. $G_i + k \leq T$); see, for example,  \citet{roth_whats_2023}.
Measuring effects according to event time is often more interpretable:  
it averages together units that have been treated for the same amount of time.
Despite this appeal, estimating effects by event time in low-rank matrix or factor model settings involves aggregating across different calendar times --- and therefore averaging across different right singular values/factors. As a result, the argument for asymptotic normality and the form of the asymptotic variance may vary from the calendar time estimand in CY.

\paragraph{Heteroskedastic-robust inference.} Ensuring that inference is robust to heteroskedasticity
is often a minimum requirement for uncertainty quantification in the social sciences, where we expect individuals or regions to be idiosyncratic.
We commend CY for extending their asymptotic results to heteroskedastic errors in the block missingness case in the supplementary materials, which goes a long way to solving this last mile problem.
A heteroskedastic-robust variance estimator in the staggered adoption case likely extends from these results and would be of great use to practitioners.

\paragraph{Hyper-parameter selection.}

CY's proposal requires choosing a tuning parameter that controls the level of 
regularization.
Hyper-parameter selection in panel data with structured patterns of missing data is a difficult problem, and various forms of cross-validation strategies have been proposed.
For example \citet{athey_matrix_2021} propose a cross-validation strategy that randomly selects subsets of the observed data so that the observed fraction is the same as the fraction of missing data in the full matrix.
A challenge is that this form of subsampling does not respect the structure of the missing data, and the average performance of the imputations may not be representative of the performance on the full matrix.
For synthetic controls with block adoption, \citet{abadie_penalized_2021} and \citet{benmichael_ascm_2021} propose a cross-validation scheme that holds out pre-treatment periods and evaluates the performance of the imputation for treated units in the held-out periods, though this may not extend to other missing data patterns.
A simple, effective, and performant approach to hyper-parameter selection would help address this last mile problem.

\paragraph{Diagnostics.}

Another key last mile challenge is the construction of diagnostic measures to assess the credibility of results.
For panel data analyses a common diagnostic is the ``event study plot'' that plots estimated event-time effects for each event time $k$. For negative event times $k < 0$, $\hat{\tau}^\text{event}_k$ serves as a placebo check. Under the parallel trends assumption, $\hat{\tau}^\text{event}_k$ should be close to zero prior to treatment, and substantial deviations from zero serve as evidence against this assumption.
Similar diagnostics are used in synthetic control analyses, where it is common to show both the event study plot and a ``gap plot'' that includes estimates of the individual treatment effects $\tau_{i(G_i + k)}$ with $k < 0$ \citep{AbadieAlbertoDiamond2010}. 
We illustrate this diagnostic for CY's proposal below.

While a useful first step, we also note that this diagnostic is likely insufficient: 
for matrix completion estimators, placebo estimates of $\hat{\tau}_{i(G_i + k)}$ near zero may reflect over-fitting.
One common and applicable alternative diagnostic is the ``in-time-placebo'' check \citep{abadie_comparative_2015}. This is an out-of-sample fit diagnostic that creates placebo treatment times $\tilde{G}_i = G_i - \ell$ for some $\ell > 0$ and estimates the treatment effects $\hat{\tau}_{i(\tilde{G}_i + k)}$ for each unit $i$ at event time $k < \ell$. Non-zero in-time-placebo estimates are an indication that some key assumption is violated.
Such diagnostics would be appropriate for CY's proposal; likely there are other useful potential diagnostics as well.

\section{Evaluating public policy with panel data}
\label{sec:policy_eval}
We conclude by discussing evaluating public policies using panel data methods.

\subsection{Challenges in evaluating policy change with panel data}
\label{sec:policy_eval_data}

A key feature of public policy evaluations is that policy changes often occur at at the jurisdictional level, e.g. at the state- or municipality-level. 
State-level policy evaluations are especially common because standardized, aggregated data at the state level is often available via federal agencies, at least for the past several decades, while coverage can be poor for more fine-grained data.
Therefore, a typical state-level policy evaluation has a limited number of units ($N\approx 50$ states) and time periods ($T\approx 50$ years), limiting the applicability of asymptotic theory that rely on growing $N$ and $T$.
This can be particularly challenging because the underlying data is often ill-conditioned: simple structures such as two-way fixed effects or one or two latent factors can explain a high degree of the variation in the data, but there is often a long tail of latent factors that account for a small but non-negligible amount of variation. For example, in the application we discuss below, the condition number of the estimated matrix is between 100-10,000, depending on the level of regularization.

Other difficulties emerge.
There is often wide variation in the timing of policy changes across states.
Early adopters serve as models for later adopters, and sometimes decades can pass between the first and last states adopting a policy.
Thus, early adopters may have relatively few pre-treatment periods, and the range in the number of pre-treatment periods across states can be large.
We also usually expect there to be treatment effect dynamics: policy effects may phase in over time as a policy is more fully rolled out, or they may appear instantly and then phase out as individuals change their behavior in response to the policy.
Finally, there is often a high degree of heteroskedasticity, as individual contexts vary widely across jurisdictions.

\subsection{A prototypical policy evaluation example}
\label{sec:example}

\begin{figure}[t]
  \centering
  \includegraphics[width=0.4\textwidth]{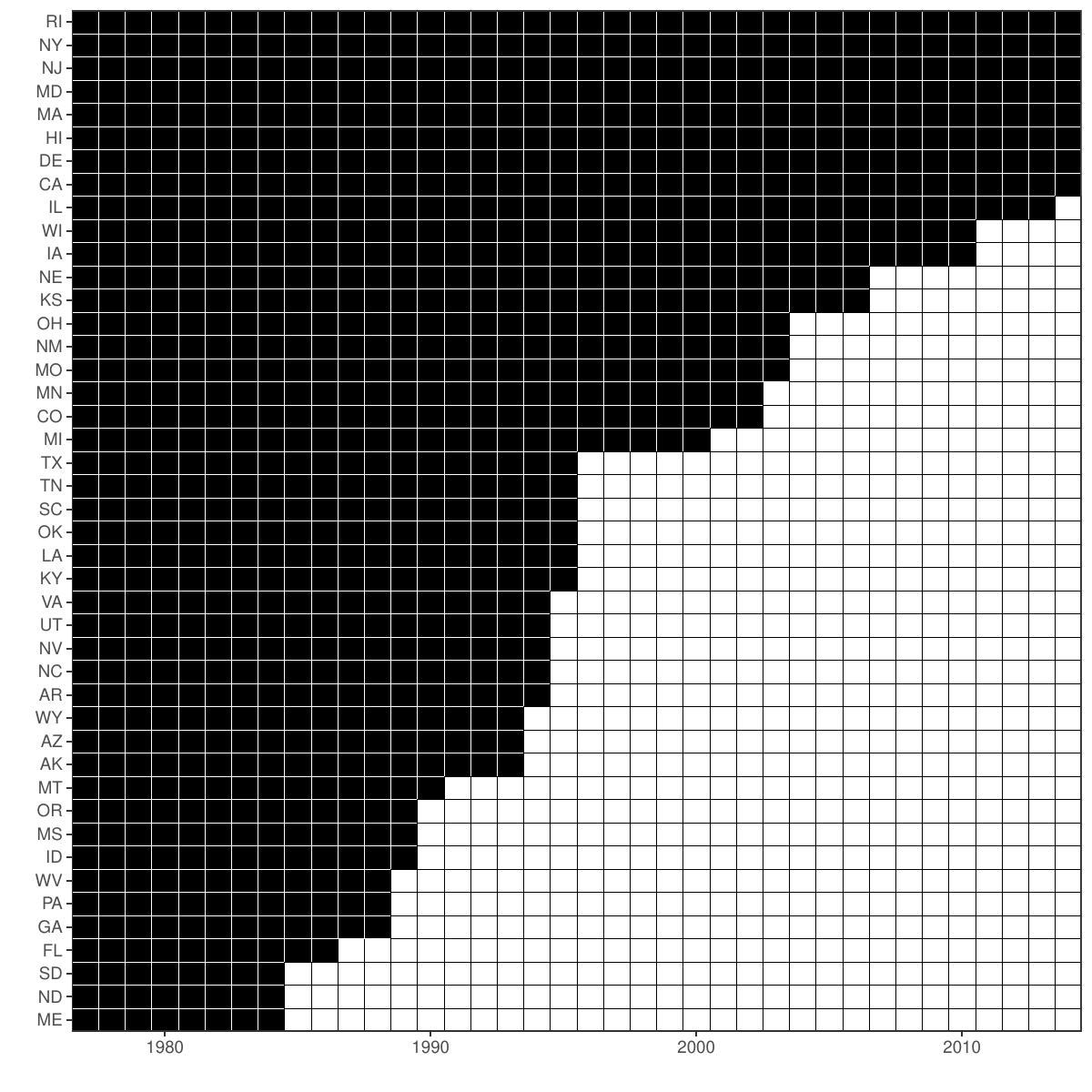}
  \caption{Treatment timing for the RTC data. Black indicates that the state had not adopted an RTC law at that time, while white indicates that it had and so $Y_{it}(\infty)$ is missing.}
  \label{fig:timing_plot}
\end{figure}

As an example policy evaluation problem, we consider the effect of ``right to carry'' (RTC) laws on violent crime.
Broadly, RTC laws allow individuals to carry concealed firearms in public either without a permit or with a permit that is easy to obtain.
Here we analyze the impact of RTC laws on violent crime following the data setup in \citet{donohue2019right}.
The panel measures the violent crime rate for each state in the United States from 1977 to 2014.\footnote{The reduced data set used for this analysis is available at \href{https://ebenmichael.github.io/assets/code/rtc_panel.csv}{\texttt{https://ebenmichael.github.io/assets/code/rtc\_panel.csv}}.}

\begin{figure}[t]
  \centering
  \includegraphics[width=0.8\textwidth]{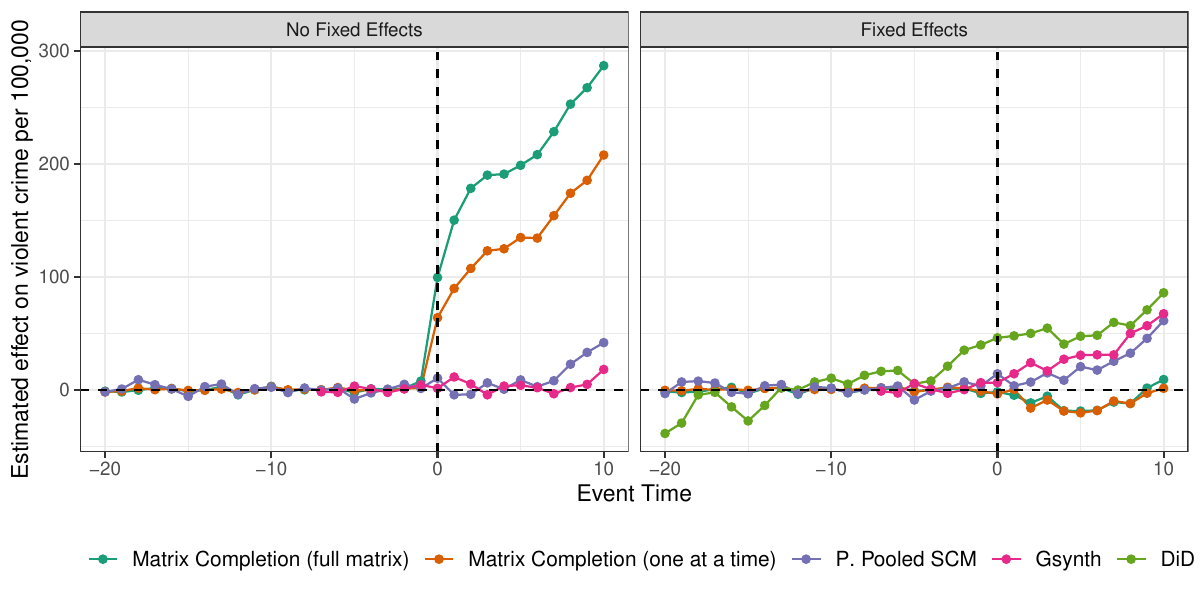}
  \caption{Estimates of the event-time effects $\tau_k^\text{event}$ for $k=-20,\ldots,10$ using (i) nuclear-norm regularized matrix completion on the entire matrix; (ii) the CY estimator; (iii) partially pooled synthetic controls \citep{benmichael_multisynth_2022}; (iv) the Gsynth estimator \citep{xu_gsynth_2017}; and (v) a DiD estimator. Estimators (i)-(iv) are estimated without (left panel) and with (right panel) residualizing out unit and time fixed effects.}
  \label{fig:estimates}
\end{figure}

Figure~\ref{fig:timing_plot} shows the timing of enacting RTC laws across the 44 states that had at least 8 years of pre-treatment observations (the remaining 6 states had enacted an RTC law before 1980).
There is a wide variation in treatment timing: the final state (IL) enacted an RTC law 29 years after the first states (ME, ND, SD).
We proceed to estimate the event-time effects of RTC laws, $\hat{\tau}_k^\text{event}$, for $k=-20,\ldots,10$ using (i) nuclear-norm regularized matrix completion on the entire matrix; (ii) the CY estimator; (iii) partially pooled synthetic controls \citep{benmichael_multisynth_2022}; and (iv) the Gsynth estimator \citep{xu_gsynth_2017}. For the matrix completion estimators we choose hyper-parameters via 5-fold cross validation on the observed unit-time pairs. Computational issues were not a concern, so we implement CY's proposal using the smallest group size possible of 1, per their suggestion.

The left panel of Figure~\ref{fig:estimates} shows the estimates. First, note that the pre-treatment estimates $\hat{\tau}_k^\text{event}$ for $k < 0$ are all close to zero for the matrix completion estimators. However, the post-treatment effect estimates for $k \geq 0$ are implausibly large: matrix completion estimates an average post-treatment increase in violent crime of 205 per 100,000 people in the years following the enactment of an RTC law.
CY's proposal gives somewhat lower estimates, with a mean post-treatment increase of 136 violent crimes per 100,000.
Again, these are implausibly large: comparing to features of the raw data, the average difference between the \emph{highest} and \emph{lowest} violent crime rates within each state during the 37 year panel period is 342 per 100,000.
Comparing to alternative estimators, we see that the matrix completion approaches estimate effects an order of magnitude larger than the alternatives (synthetic control average: 11.3 per 100,000; Gsynth average: 3.86 per 100,000).

Finally, we can also consider using two-way fixed effects as a \emph{pre-processing} step by estimating unit and time fixed effects from the observed data matrix and then subtracting them from the observed violent crime rate before using the various estimators. Such pre-processing has proven effective in synthetic control settings because it directly accounts for level differences across units and time \citep[e.g.][]{ferman_imperfect_2021}; if no additional estimation step is applied, this yields a standard DiD estimator.
The right panel of Figure~\ref{fig:estimates} shows the corresponding event study plot on residualized data,
along with an additional DiD estimate. 
The unadjusted DiD estimate shows clear evidence against parallel trends, with pre-treatment placebo estimates far from zero. 
Nonetheless, after removing unit and time fixed effects  
as a pre-processing step, the matrix completion estimates are in much greater alignment with alternative approaches. Matrix completion on the full matrix and the CY estimator estimate a mean post-treatment decrease of 8.5 and 10.1 violent crimes per 100,000, respectively, compared to an estimated increase of 23 and 32 violent crimes per 100,000 for the synthetic control and Gsynth estimators, respectively.
A bootstrap-based approximate 95\% confidence interval for the synthetic control estimate ranges from -48 to 116, a wide range that includes the matrix completion estimates with fixed effects pre-processing, but not without pre-processing.

\clearpage
\pdfbookmark[1]{References}{References}
\singlespacing
\bibliographystyle{apalike}
\bibliography{citations}

@article{wickham_split-apply-combine_2011,
	title = {The {Split}-{Apply}-{Combine} {Strategy} for {Data} {Analysis}},
	volume = {40},
	copyright = {Copyright (c) 2009 Hadley  Wickham},
	issn = {1548-7660},
	url = {https://doi.org/10.18637/jss.v040.i01},
	doi = {10.18637/jss.v040.i01},
	language = {en},
	urldate = {2025-07-22},
	journal = {Journal of Statistical Software},
	author = {Wickham, Hadley},
	month = apr,
	year = {2011},
	pages = {1--29},
}

@article{sun_did_2021,
	title = {Estimating dynamic treatment effects in event studies with heterogeneous treatment effects},
	volume = {225},
	issn = {03044076},
	url = {https://linkinghub.elsevier.com/retrieve/pii/S030440762030378X},
	doi = {10.1016/j.jeconom.2020.09.006},
	language = {en},
	number = {2},
	urldate = {2024-02-05},
	journal = {Journal of Econometrics},
	author = {Sun, Liyang and Abraham, Sarah},
	month = dec,
	year = {2021},
	pages = {175--199},
}

@article{callaway_did_2021,
	title = {Difference-in-{Differences} with {Multiple} {Time} {Periods}},
	volume = {225},
	number = {2},
	journal = {Journal of Econometrics},
	author = {Callaway, Brantly and Sant'Anna, Pedro H.C.},
	year = {2021},
	pages = {200--230},
}

@article{athey2018design,
	title = {Design-based analysis in {Difference}-{In}-{Differences} settings with staggered adoption},
	volume = {226},
	issn = {18726895},
	url = {https://doi.org/10.1016/j.jeconom.2020.10.012},
	doi = {10.1016/j.jeconom.2020.10.012},
	number = {1},
	journal = {Journal of Econometrics},
	author = {Athey, Susan and Imbens, Guido W.},
	year = {2022},
	pages = {62--79},
}

@article{AbadieAlbertoDiamond2010,
	title = {Synthetic {Control} {Methods} for {Comparative} {Case} {Studies}: {Estimating} the {Effect} of {California}’s {Tobacco} {Control} {Program}},
	volume = {105},
	issn = {0162-1459},
	url = {http://amstat.tandfonline.com/action/journalInformation?journalCode=uasa20},
	doi = {10.1198/jasa.2009.ap08746},
	number = {490},
	urldate = {2017-10-18},
	journal = {Journal of the American Statistical Association},
	author = {Abadie, Alberto and Diamond, Alexis and Hainmueller, Jens},
	year = {2010},
	pmid = {741578133},
	note = {ISBN: 0162-1459},
	pages = {493--505}
}

@article{goodman-bacon_did_2021,
	title = {Difference-in-differences with variation in treatment timing},
	volume = {225},
	issn = {0304-4076},
	doi = {10.1016/j.jeconom.2021.03.014},
	number = {2},
	journal = {Journal of Econometrics},
	author = {Goodman-Bacon, Andrew},
	year = {2021},
	pages = {254--277},
}

@article{borusyak_revisiting_2024,
	title = {Revisiting {Event}-{Study} {Designs}: {Robust} and {Efficient} {Estimation}},
	volume = {91},
	issn = {0034-6527},
	shorttitle = {Revisiting {Event}-{Study} {Designs}},
	doi = {10.1093/restud/rdae007},
	number = {6},
	journal = {The Review of Economic Studies},
	author = {Borusyak, Kirill and Jaravel, Xavier and Spiess, Jann},
	year = {2024},
	pages = {3253--3285}
}

@article{roth_whats_2023,
	title = {What's trending in difference-in-differences? {A} synthesis of the recent econometrics literature},
	volume = {235},
	issn = {0304-4076},
	shorttitle = {What's trending in difference-in-differences?},
	doi = {10.1016/j.jeconom.2023.03.008},
	number = {2},
	urldate = {2024-08-15},
	journal = {Journal of Econometrics},
	author = {Roth, Jonathan and Sant'Anna, Pedro H. C. and Bilinski, Alyssa and Poe, John},
	month = aug,
	year = {2023},
	pages = {2218--2244},
}

@article{benmichael_multisynth_2022,
	title = {Synthetic controls with staggered adoption},
	volume = {84},
	issn = {14679868},
	doi = {10.1111/rssb.12448},
	number = {2},
	journal = {Journal of the Royal Statistical Society. Series B: Statistical Methodology},
	author = {Ben-Michael, Eli and Feller, Avi and Rothstein, Jesse},
	year = {2022},
	pages = {351--381},
}

@article{abadie_penalized_2021,
	title = {A {Penalized} {Synthetic} {Control} {Estimator} for {Disaggregated} {Data}},
	volume = {116},
	issn = {1537274X},
	url = {https://doi.org/10.1080/01621459.2021.1971535},
	doi = {10.1080/01621459.2021.1971535},
	number = {536},
	journal = {Journal of the American Statistical Association},
	author = {Abadie, Alberto and L'Hour, Jérémy},
	year = {2021},
	note = {Publisher: Taylor \& Francis},
	pages = {1817--1834},
}

@article{zubizarreta_matching_2011,
	title = {Matching for {Several} {Sparse} {Nominal} {Variables} in a {Case}-{Control} {Study} of {Readmission} {Following} {Surgery}},
	volume = {65},
	issn = {0003-1305},
	url = {https://doi.org/10.1198/tas.2011.11072},
	doi = {10.1198/tas.2011.11072},
	number = {4},
	urldate = {2025-07-24},
	journal = {The American Statistician},
	author = {Zubizarreta, José R. and Reinke, Caroline E. and Kelz, Rachel R. and Silber, Jeffrey H. and Rosenbaum, Paul R.},
	month = nov,
	year = {2011},
	pmid = {25418991},
	pages = {229--238}
}

@article{Kreif2016,
	title = {Estimating causal effects: considering three alternatives to difference-in-differences estimation},
	volume = {16},
	doi = {10.1007/s10742-016-0146-8},
	journal = {Health Services and Outcomes Research Methodology},
	author = {Kreif, Noémi and Grieve, Richard and Sutton, Matthew and Sekhon, Jasjeet},
	year = {2016},
	pages = {1--21},
}

@article{Robbins2017,
	title = {A {Framework} for {Synthetic} {Control} {Methods} {With} {High}-{Dimensional}, {Micro}-{Level} {Data}: {Evaluating} a {Neighborhood}-{Specific} {Crime} {Intervention}},
	volume = {112},
	issn = {1537274X},
	doi = {10.1080/01621459.2016.1213634},
	number = {517},
	journal = {Journal of the American Statistical Association},
	author = {Robbins, Michael W. and Saunders, Jessica and Kilmer, Beau},
	year = {2017},
	pages = {109--126}
}

@article{athey_matrix_2021,
	title = {Matrix {Completion} {Methods} for {Causal} {Panel} {Data} {Models}},
	volume = {116},
	issn = {1537274X},
	doi = {10.1080/01621459.2021.1891924},
	number = {536},
	journal = {Journal of the American Statistical Association},
	author = {Athey, Susan and Bayati, Mohsen and Doudchenko, Nikolay and Imbens, Guido and Khosravi, Khashayar},
	year = {2021},
	pages = {1716--1730}
}

@article{benmichael_ascm_2021,
	title = {The {Augmented} {Synthetic} {Control} {Method}},
	volume = {116},
	issn = {1537274X},
	doi = {10.1080/01621459.2021.1929245},
	number = {536},
	journal = {Journal of the American Statistical Association},
	author = {Ben-Michael, Eli and Feller, Avi and Rothstein, Jesse},
	year = {2021},
	pages = {1789--1803},
}

@article{abadie_comparative_2015,
	title = {Comparative {Politics} and the {Synthetic} {Control} {Method}},
	volume = {59},
	issn = {15405907},
	doi = {10.1111/ajps.12116},
	number = {2},
	journal = {American Journal of Political Science},
	author = {Abadie, Alberto and Diamond, Alexis and Hainmueller, Jens},
	year = {2015},
	pages = {495--510},
}

@article{donohue2019right,
  title={Right-to-Carry Laws and Violent Crime: A Comprehensive Assessment Using Panel Data and a State-Level Synthetic Control Analysis},
  author={Donohue, John J and Aneja, Abhay and Weber, Kyle D},
  journal={Journal of Empirical Legal Studies},
  volume={16},
  number={2},
  pages={198--247},
  year={2019},
  publisher={Wiley Online Library}
}

@article{xu_gsynth_2017,
	title = {Generalized {Synthetic} {Control} {Method}: {Causal} {Inference} with {Interactive} {Fixed} {Effects} {Models}},
	volume = {25},
	issn = {1047-1987, 1476-4989},
	shorttitle = {Generalized {Synthetic} {Control} {Method}},
	url = {https://www.cambridge.org/core/journals/political-analysis/article/generalized-synthetic-control-method-causal-inference-with-interactive-fixed-effects-models/B63A8BD7C239DD4141C67DA10CD0E4F3},
	doi = {10.1017/pan.2016.2},
	number = {1},
	urldate = {2025-07-25},
	journal = {Political Analysis},
	author = {Xu, Yiqing},
	month = jan,
	year = {2017},
	pages = {57--76}
}

@article{ferman_imperfect_2021,
	title = {Synthetic controls with imperfect pretreatment fit},
	volume = {12},
	copyright = {Copyright © 2021 The Authors.},
	issn = {1759-7331},
	url = {https://onlinelibrary.wiley.com/doi/abs/10.3982/QE1596},
	doi = {10.3982/QE1596},
	number = {4},
	urldate = {2025-01-17},
	journal = {Quantitative Economics},
	author = {Ferman, Bruno and Pinto, Cristine},
	year = {2021},
	pages = {1197--1221}
}

\end{document}